\documentclass[aps,prd,reprint,preprintnumbers,superscriptaddress,showpacs,twocolumn,nofootinbib]{revtex4-1}
\usepackage{latexsym,graphicx,amssymb,amsmath,mathrsfs}
\usepackage{setspace,bm}
\usepackage[colorlinks=true, pdfstartview=FitV, linkcolor=red, citecolor=blue, urlcolor=blue]{hyperref}
\usepackage[caption=false]{subfig}
\usepackage{color}

\graphicspath{{./Figures/}}

\newcommand{\diff}{\mathrm{d}}
\newcommand{\p}{\partial}
\newcommand{\ve}{\varepsilon}
\newcommand{\Diff}{{\mathcal{D}}}

\newcommand{\be}{\begin{equation}}      
\newcommand{\ee}{\end{equation}}      
\newcommand{\bea}{\begin{eqnarray}}      
\newcommand{\eea}{\end{eqnarray}}

\newcommand{\tr}{\mathrm{tr}}

\newcommand{\im}{\mathrm{i}}

\newcommand{\calA}{\mathcal{A}}

\newcommand{\rmc}{\mathrm{c}}
\newcommand{\rmf}{\mathrm{f}}
\newcommand{\rmq}{\mathrm{q}}
\newcommand{\rmcq}{\mathrm{c\mbox{-}q}}
\newcommand{\rmfq}{\mathrm{f\mbox{-}q}}

\usepackage[normalem]{ulem}

\begin{document}

\title{Anomaly matching for phase diagram of massless $\mathbb{Z}_N$-QCD}

\author{Yuya Tanizaki}
\email{yuya.tanizaki@riken.jp}
\affiliation{RIKEN BNL Research Center, Brookhaven National Laboratory, Upton, NY 11973, USA}

\author{Yuta Kikuchi}
\email{kikuchi@ruby.scphys.kyoto-u.ac.jp}
\affiliation{Department of Physics, Kyoto University, Kyoto 606-8502, Japan}
\affiliation{Department of Physics and Astronomy, Stony Brook University, Stony Brook, New York 11794-3800, USA}

\author{Tatsuhiro Misumi}
\email{misumi@phys.akita-u.ac.jp}
\affiliation{Department of Mathematical Science, Akita University, Akita 010-8502, Japan}
\affiliation{Research and Education Center for Natural Sciences, Keio University, Kanagawa 223-8521, Japan}
\affiliation{iTHEMS, RIKEN, Wako, Saitama 351-0198, Japan}

\author{Norisuke Sakai}
\email{norisuke.sakai@gmail.com}
\affiliation{Research and Education Center for Natural Sciences, Keio University, Kanagawa 223-8521, Japan}
\affiliation{iTHEMS, RIKEN, Wako, Saitama 351-0198, Japan}

\date{\today}

\begin{abstract}
We elucidate that the phase diagram of massless $N$-flavor QCD under $\mathbb{Z}_N$ flavor-twisted boundary condition (massless $\mathbb{Z}_N$-QCD) is constrained by an 't Hooft anomaly involving two-form gauge fields. 
As a consequence, massless $\mathbb{Z}_N$-QCD turns out to realize persistent order at any temperatures and quark chemical potentials, namely, the symmetric and gapped phase is strictly forbidden. 
This is the first result on the finite-$(T,\mu)$ phase diagram in QCD-type theories based on anomaly matching related to center and discrete axial symmetries.
\end{abstract}


\maketitle

\section{Introduction}\label{sec:intro}
Conventional wisdom of classical statistical physics tells us that different phases of matter are distinguished by different patterns of spontaneous symmetry breaking (SSB).  
This idea develops Ginzburg-Landau effective theory~\cite{landau1937theory,ginzburg1950theory} and 
gives a useful guideline to understand nature of various matters. 
For quantum field theories (QFTs), however, the Ginzburg-Landau description is not always a good starting point especially when QFTs have an 't~Hooft anomaly~\cite{tHooft:1979rat, Frishman:1980dq, Coleman:1982yg}. This is because the anomaly matching argument rules out the symmetric and gapped phase, which we call the trivial phase. 
Necessity of beyond-Ginzburg-Landau description has also been clarified, for instance, by Haldane conjecture~\cite{Haldane:1982rj, Haldane:1983ru, Affleck:1986pq, Affleck:1987vf, Haldane:1988zza}, deconfined quantum criticality~\cite{Senthil:2004aza, PhysRevB.70.144407, Wang:2017txt}, in condensed matter physics. 
Recent development about QFTs suggests that the absence of trivial phase is closely related to the existence of 't Hooft anomaly, and many quantum systems are reconsidered from this viewpoint~\cite{Vishwanath:2012tq, Kapustin:2014lwa, Kapustin:2014zva, Cho:2014jfa, Wang:2014pma, Witten:2015aba,Seiberg:2016rsg,Tachikawa:2016cha, Tachikawa:2016nmo, Gaiotto:2017yup,Tanizaki:2017bam, Komargodski:2017dmc, Komargodski:2017smk, Cho:2017fgz,Shimizu:2017asf, Wang:2017loc, Metlitski:2017fmd,Kikuchi:2017pcp, Gaiotto:2017tne, Tanizaki:2017qhf, Yamazaki:2017dra, Cherman:2017dwt, Guo:2017xex}.

In this paper, we claim that quantum chromodynamics (QCD) should also be added to the list of quantum systems that require the beyond-Ginzburg-Landau treatment. 
QCD is the fundamental law of strong interaction, thus it can provide the first-principle calculation of finite-density nuclear matters, which is an important subject to understand, for instance, the interior of neutron stars~\cite{Alford:2007xm, Fukushima:2010bq, Masuda:2012kf}. Because of the notorious sign problem, however, no \textit{ab initio} lattice QCD simulation is yet available at finite quark chemical potentials~\cite{Barbour:1997ej, Karsch:2001cy, Muroya:2003qs, Cohen:2003kd, Tanizaki:2015rda}.

We consider massless $N$-flavor QCD with color $N$ in four dimensions, which has the vector-like continuous symmetry and discrete axial symmetry, and we show by an explicit calculation that there is a mixed 't~Hooft anomaly between those symmetries. 
Imposing the flavor-twisted boundary condition in the compactified direction so that the $\mathbb{Z}_N$ center symmetry appears (as known as $\mathbb{Z}_N$-QCD~\cite{Tanizaki:2017qhf, Kouno:2012zz, Sakai:2012ika, Kouno:2013zr, Kouno:2013mma, Poppitz:2013zqa, Iritani:2015ara, Kouno:2015sja, Hirakida:2016rqd, Hirakida:2017bye, Cherman:2017tey}), the above mixed anomaly turns out to survive at any temperatures and quark chemical potentials.  
Anomaly matching precludes existence of the trivial phase for massless $\mathbb{Z}_N$-QCD. 
We discuss consistency of our result with previous studies, and argue its implications to the phase diagram of massless $\mathbb{Z}_N$-QCD.

\section{Anomaly matching of massless $\mathbb{Z}_N$-QCD}

We first compute an 't Hooft anomaly of massless $N$-flavor QCD in four dimensions, and derive the anomaly of massless $\mathbb{Z}_N$-QCD using the technique given in \cite{Tanizaki:2017qhf}. 

\paragraph{Massless $N$-flavor QCD and 't Hooft anomaly}
We consider four-dimensional $SU(N)$ Yang-Mills theory with $N$ massless Dirac fermions in the fundamental representation, i.e. massless QCD with $N_\rmc=N_\rmf=N$. 
We discriminate the color and flavor $SU(N)$ groups by denoting them as $SU(N)_\rmc$ and $SU(N)_\rmf$, respectively. 
$SU(N)_\rmc$ gauge field is denoted as $a$, and the field strength is given by $G_{\rmc}=\diff a+\im a\wedge a$. 
We represent the Dirac field $q=(q_{cf})_{c,f=1,\ldots,N}$ as an $N\times N$ matrix-valued Grassmannian variable, on which $SU(N)_\rmc\times SU(N)_{\rmf}$ acts as a bifundamental representation: $q\mapsto U_\rmc q U_\rmf^{\dagger}$ for $(U_\rmc,U_\rmf)\in SU(N)_\rmc\times SU(N)_\rmf$. 
The classical action of this theory is given by 
\be
S={1\over 2g^2}\int \tr_\rmc(G_{\mathrm{c}}\wedge *G_{\mathrm{c}})+\int \diff^4 x\, \tr_\rmf\left\{\overline{q}\gamma_{\mu}D_{\mu}(a)q\right\}, 
\label{eq:massless_qcd}
\ee
where $\tr_\rmc$ and $\tr_\rmf$ denote the trace operation in color and flavor spaces, respectively, and $D(a)=\diff+\im a$ is the covariant derivative. 

We pay attention to the vector-like symmetry,
\be
{SU(N)_\rmf\times U(1)_\rmq\over (\mathbb{Z}_N)_{\rmcq}\times (\mathbb{Z}_N)_{\rmfq}},
\label{eq:vector_symmetry}
\ee
and the anomaly free subgroup of $U(1)$ axial symmetry, $(\mathbb{Z}_{2N})_{\mathrm{axial}}$. 
We shall explain details of these symmetries. 

The quark field $q$ is in the representation of vector-like symmetry $SU(N)_\rmc\times SU(N)_\rmf\times U(1)_\rmq$, $q\mapsto \mathrm{e}^{\im \alpha}U_\rmc q U_\rmf^\dagger$ for $(U_\rmc,U_\rmf,\mathrm{e}^{\im\alpha})\in SU(N)_\rmc\times SU(N)_\rmf\times U(1)_\rmq$. However, there are two $\mathbb{Z}_N$ subgroups that do not change $q$. 
One is generated by $(\omega,1,\omega^{-1})\in SU(N)_\rmc\times SU(N)_\rmf\times U(1)_\rmq$ ($\omega=\mathrm{e}^{2\pi\im/N}$), and we denote it as $(\mathbb{Z}_N)_{\rmcq}$.  
The another is generated by $(1,\omega,\omega)\in SU(N)_\rmc\times SU(N)_\rmf\times U(1)_\rmq$, and we denote it as $(\mathbb{Z}_N)_{\rmfq}$.  
We thus regard (\ref{eq:vector_symmetry}) as a symmetry with faithful representations. 

For massless quark field $q$, there is the axial $U(1)$ symmetry, $q\mapsto \mathrm{e}^{\im\alpha \gamma_5}q$, for the classical action (\ref{eq:massless_qcd}), but the fermion functional integration measure $\Diff\overline{q}\Diff q$ is changed as
$\Diff \overline{q}\Diff q\mapsto \Diff \overline{q}\Diff q\exp\left(\im 2\alpha{N\over 8\pi^2}\int \tr_\rmc(G_c\wedge G_c)\right)$~\cite{Fujikawa:1979ay,Fujikawa:1980eg}. 
Only if $\alpha$ is quantized to $2\pi/(2N)$, does this transformation become symmetry of $N$-flavor massless QCD. This is the $(Z_{2N})_{\mathrm{axial}}$ symmetry. 

In order to detect the mixed 't Hooft anomaly between the vector-like symmetry \eqref{eq:vector_symmetry} and $(\mathbb{Z}_{2N})_{\mathrm{axial}}$, we introduce the background gauge fields for the vector-like symmetry. 
Background gauge fields consist of following ingredients:
\begin{itemize}
\item $SU(N)_\rmf$ one-form gauge field: $A_\rmf$ 
\item $U(1)_\rmq$ one-form gauge field: $A_\rmq$
\item $(\mathbb{Z}_N)_{\rmcq}$ two-form gauge field: $B_{\rmc}$ ($N B_{\rmc}=\diff C_{\rmc}$)
\item $(\mathbb{Z}_N)_{\rmfq}$ two-form gauge field: $B_{\rmf}$ ($N B_{\rmf}=\diff C_{\rmf}$)
\end{itemize}
$\mathbb{Z}_N$ two-form gauge field $B$ may require an explanation: We realize it as a pair of $U(1)$ two-form gauge field $B$ and $U(1)$ one-form gauge field $C$ with the constraint $NB=\diff C$. 
It generates a $\mathbb{Z}_N$ one-form symmetry transformation acting on Wilson line operators~\cite{Gaiotto:2014kfa}, which arises after gauging $SU(N)_\rmf\times U(1)_\rmq$ \cite{Tanizaki:2017bam, Shimizu:2017asf, Gaiotto:2017tne, Tanizaki:2017qhf}.
Introducing the two-form gauge fields $B_{\rmc}$ and $B_{\rmf}$, the $SU(N)_{\rmc,\rmf}$ gauge connections are once promoted to $U(N)$ gauge connections~\cite{Kapustin:2014gua},
\be
\widetilde{a}=a+{1\over N}C_{\rmc},\;\; \widetilde{A_{\rmf}}=A_{\rmf}+{1\over N}C_{\rmf}.
\ee
The covariant derivative is replaced as 
\be
D(\widetilde{a},\widetilde{A_{\rmf}},A_{\rmq})q=(\diff + \im \widetilde{a}-\im (\widetilde{A_\rmf})^t+\im A_{\rmq})q,
\label{eq:covariant_derivative}
\ee
by the minimal coupling procedure. More explicitly, 
\be
(D_{\mu}q)_{cf}=(\p_{\mu}+\im (A_\rmq)_{\mu}) q_{cf}+ \im \widetilde{a}_{\mu,cc'}q_{c'f}-\im q_{cf'}(\widetilde{A_\rmf})_{\mu,f'f}. 
\label{eq:covariant_derivative_elements}
\ee
(\ref{eq:covariant_derivative}) must be regarded as a short-hand notation of (\ref{eq:covariant_derivative_elements}), where the repeated indices are summed.

With this setup, we consider the partition function under these background gauge fields, $\mathcal{Z}[(A_{\rmf},A_{\rmq},B_{\rmc},B_{\rmf})]$. Applying the $(\mathbb{Z}_{2N})_{\mathrm{axial}}$ transformation, we can compute the change of the partition function,
\be
\mathcal{Z}[(A_{\rmf},A_{\rmq},B_{\rmc},B_{\rmf})]\mapsto \mathcal{Z}[(A_{\rmf},A_{\rmq},B_{\rmc},B_{\rmf})]\exp(\im \mathcal{A}). 
\ee
The 't Hooft anomaly $\mathcal{A}$ again comes from the change of the fermion measure $\Diff\overline{q}\Diff q$, and Fujikawa method gives
\be
\label{eq:anomaly_Fujikawa_background}
\mathcal{A}=2{2\pi\over 2N}{1\over 8\pi^2}\int \tr_{\rmc,\rmf}\left[F\wedge F\right], 
\ee
with $F=\im^{-1} D(\widetilde{a},\widetilde{A_{\rmf}},A_{\rmq})\wedge D(\widetilde{a},\widetilde{A_{\rmf}},A_{\rmq})$. $\tr_{\rmc,\rmf}$ denotes the trace operation over both color and flavor spaces. 
The explicit computation shows that 
\be
\label{eq:field_strength_background}
F=(\diff \widetilde{a} +\im \widetilde{a}\wedge \widetilde{a})-(\diff \widetilde{A_\rmf}+\im \widetilde{A_\rmf}\wedge \widetilde{A_\rmf})^t+\diff A_{\rmq}.
\ee
We obtain from (\ref{eq:anomaly_Fujikawa_background}) and (\ref{eq:field_strength_background}) that (modulo $2\pi$)
\be
\calA
=-{N\over 2\pi}\int B_{\rmc}\wedge B_{\rmf}. 
\label{eq:anomaly_massless4d_qcd}
\ee
Since $\calA$ is nontrivial, there is a mixed 't Hooft anomaly between the vector-like symmetry \eqref{eq:vector_symmetry} and $(\mathbb{Z}_{2N})_{\mathrm{axial}}$ for four-dimensional massless $N$-flavor QCD. 

\paragraph{Massless $\mathbb{Z}_N$-QCD and 't Hooft anomaly}

We perform the circle compactification of size $L$ and put the boundary condition on the quark field as 
\be
q(\bm{x},x_4+L)=q(\bm{x},x_4)\Omega,
\label{eq:TBC}
\ee
where ($\omega=\mathrm{e}^{2\pi\im/N}$, $-\pi<\phi\le \pi$, and $\mu\in\mathbb{R}$)
\be
\Omega=\mathrm{diag}[1,\omega,\ldots,\omega^{N-1}]\mathrm{e}^{\im \phi+\mu L}. 
\ee
Here, $\mu$ is the quark chemical potential. This theory is called $\mathbb{Z}_N$-QCD and denote its partition function as $\mathcal{Z}_{\Omega}$~\cite{Kouno:2012zz, Sakai:2012ika, Kouno:2013zr, Kouno:2013mma, Poppitz:2013zqa, Iritani:2015ara, Kouno:2015sja, Hirakida:2016rqd, Hirakida:2017bye, Cherman:2017tey}. Although the twisted boundary condition is imposed, we call $T\equiv L^{-1}$ a temperature of massless $\mathbb{Z}_N$-QCD. 
By performing the gauge transformation, we can express this symmetry-twisted partition function as 
\be
\mathcal{Z}_{\Omega}=\mathrm{tr}_{\mathcal{H}}\left[\mathrm{e}^{-L(\widehat{H}-\mu \widehat{Q})}\exp\left(\im \sum_{f=1}^{N}{2\pi f\over N}\widehat{Q}_f\right)\right], 
\label{eq:operator_expression}
\ee
where $\widehat{H}$ is the QCD Hamiltonian, $\mathcal{H}$ is the QCD Hilbert space, $\widehat{Q}_f=\int \diff^3 \bm{x}\, \widehat{q}^{\dagger}_f \widehat{q}_f(\bm{x})$ is the quark number of flavor $f$, and $\widehat{Q}=\sum_f \widehat{Q}_f$ is the total quark number (For simplicity of the expression, we set $\phi=\pi$). 
This twisted boundary condition plays an essential role for the existence of 't~Hooft anomaly in the $S^1$-compactified theory~\cite{Tanizaki:2017qhf}. 
The same boundary condition is also important for the large-$N$ volume independence of $\mathbb{C}P^{N-1}$ model~\cite{Tanizaki:2017qhf, Dunne:2012ae, Dunne:2012zk, Misumi:2014jua, Misumi:2016fno, Sulejmanpasic:2016llc}. 

We first describe the symmetry of massless $\mathbb{Z}_N$-QCD as three-dimensional QFT. 
There are three symmetries of importance, $(\mathbb{Z}_N)_{\mathrm{shift},\rmcq}$, $(U(1)^{N-1}_\rmf\times U(1)_\rmq)/ ((\mathbb{Z}_N)_{\rmcq}\times (\mathbb{Z}_N)_{\rmfq})$, and $(\mathbb{Z}_{2N})_{\mathrm{axial}}$. 
To explain $(\mathbb{Z}_N)_{\mathrm{shift},\rmcq}$, it is better to regard $\Omega$ as the background holonomy of vector-like symmetry along the compactified direction,
\be
\Omega=\left\{\mathcal{P}\exp\left(\im\int_0^L A_\rmf\right)\right\}^{-1} \exp\left(\im \int_0^L A_\rmq\right). 
\ee
We consider the $(\mathbb{Z}_N)_{\mathrm{shift}}$ transformation defined by 
\be
q=(q_{c,1},\ldots,q_{c,N})\mapsto (q_{c,2},\ldots,q_{c,N},q_{c,1})=:q S^{-1}.
\ee
Under this transformation, the boundary condition (\ref{eq:TBC}) is changed as 
\be
\Omega\mapsto S\Omega S^{-1}=\omega \Omega. 
\ee
To maintain the boundary condition, we combine the $(\mathbb{Z}_N)_{\rmcq}$ zero-form and $(\mathbb{Z}_N)_{\mathrm{shift}}$ transformations, and we define the $(\mathbb{Z}_N)_{\mathrm{shift},\rmcq}$ symmetry~\cite{Tanizaki:2017qhf,Cherman:2017tey},
\be
\tr_\rmc(\Phi)\mapsto \omega\, \tr_\rmc(\Phi),\, q\mapsto q S^{-1},  
\ee
where $\Phi:=\mathcal{P}\exp\left(\im\int_{S^1}a\right)$ is the color Polyakov loop. 
The vector-like symmetry must commute with the matrix $\Omega$ defining the boundary condition, and the symmetry (\ref{eq:vector_symmetry}) is explicitly broken to its maximal Abelian subgroup, $(U(1)^{N-1}_\rmf\times U(1)_\rmq)/ ((\mathbb{Z}_N)_{\rmcq}\times (\mathbb{Z}_N)_{\rmfq})$. 

Introducing the three-dimensional background gauge fields for $(\mathbb{Z}_N)_{\mathrm{shift},\rmcq}$ and $(U(1)^{N-1}_\rmf\times U(1)_\rmq)/ ((\mathbb{Z}_N)_{\rmcq}\times (\mathbb{Z}_N)_{\rmfq})$, we obtain the following two-form gauge fields by use of the four-dimensional language~\cite{Tanizaki:2017qhf},
\be
B_\rmc=B^{(1)}_{\rmc}\wedge L^{-1}\diff x^4+B_\rmc^{(2)}, \; \; B_\rmf=B_\rmf^{(2)}. 
\ee
Here, $B^{(1)}_{\rmc}$ is a one-form gauge field for $(\mathbb{Z}_N)_{\mathrm{shift},\rmcq}$ zero-form symmetry in three dimensions, and $B_{\rmc,\rmf}^{(2)}$ are two-form gauge fields for $(\mathbb{Z}_N)_{\rmcq,\rmfq}$ one-form symmetries, respectively, also in three dimensions. 
Substituting it into the four-dimensional anomaly (\ref{eq:anomaly_massless4d_qcd}), we obtain the anomaly for massless $\mathbb{Z}_N$-QCD: 
\be
\calA=-{N\over 2\pi}\int B_\rmc^{(1)}\wedge B_\rmf^{(2)} \in {2\pi\over N}\mathbb{Z}. 
\label{eq:tHooft_anomaly_ZnQCD}
\ee
This gives the mixed 't Hooft anomaly of massless $\mathbb{Z}_N$-QCD among $(\mathbb{Z}_N)_{\mathrm{shift},\rmcq}$, $(U(1)^{N-1}_\rmf\times U(1)_\rmq)/ ((\mathbb{Z}_N)_{\rmcq}\times (\mathbb{Z}_N)_{\rmfq})$, and $(\mathbb{Z}_{2N})_{\mathrm{axial}}$, for any $N\ge 2$. 
To match the anomaly, phase diagram of massless $\mathbb{Z}_N$-QCD must realize the persistent order at any temperatures $L^{-1}$ and quark chemical potentials $\mu$: The trivial phase is strictly excluded by the anomaly matching condition.
%
%

\section{Phase structure of massless $\mathbb{Z}_N$-QCD}

In certain limits of $T=L^{-1}$ and $\mu$, the phase structure of massless $\mathbb{Z}_N$-QCD is calculable, and thus we can check how the anomaly matching is satisfied in those limits. 
We consider high temperature or large chemical potential as a limit where the reliable perturbative calculations are available. 
We also note that the lattice simulation~\cite{Iritani:2015ara} suggests that at large-$L$ and small-$\mu$ region the anomaly (\ref{eq:tHooft_anomaly_ZnQCD}) is matched by SSB of $(\mathbb{Z}_{2N})_{\mathrm{axial}}\to \mathbb{Z}_2$. 
Since this naturally breaks the continuous chiral symmetry by developing quark bilinear condensate $\langle \overline{q}q\rangle$, this is also consistent with an 't Hooft anomaly of continuous chiral symmetry~\cite{tHooft:1979rat} at $L=\infty$ and $\mu=0$. 

\paragraph{High temperature limit}
At small $L$ and $\mu=0$, we compute the one-loop effective potential of the color Polyakov loop $\Phi$ \cite{Gross:1980br, Weiss:1980rj}. 
Gluon contribution is 
\be
V_{\mathrm{gl}}(\Phi)=-{2\over \pi^2 L^4}\sum_{n\ge 1}{1\over n^4}\left(|\tr_\rmc \left(\Phi^n\right)|^{2}-1\right). 
\ee
Quark contribution (at $\mu=0$) is 
\be
V_{\mathrm{qk}}(\Phi)
={2 N^{-3}\over \pi^2 L^4}\sum_{n\ge 1}{1\over n^4}\left(\mathrm{e}^{\im n N \phi}\tr_\rmc(\Phi^{N n})+\mathrm{h.c.}\right).  
\ee
Gluon contribution is $V_{\mathrm{gl}}=O(N^2)$, while the quark contribution is $V_{\mathrm{qk}}=O(N^{-2})$~\cite{Cherman:2017tey}. Therefore, the high-temperature behavior of $\mathbb{Z}_N$-QCD is essentially determined by the gluonic contribution. 
$(\mathbb{Z}_N)_{\mathrm{shift},\rmcq}$ is spontaneously broken, and the vacuum is $\Phi\propto \bm{1}_N$.  
The 't~Hooft anomaly (\ref{eq:tHooft_anomaly_ZnQCD}) is matched by SSB of $(\mathbb{Z}_N)_{\mathrm{shift},\rmcq}$. 
Strictly speaking, this perturbative argument is subtle because the three-dimensional effective theory is a strongly-coupled gauge theory, but the above observation is consistent with lattice simulation~\cite{Iritani:2015ara}. 

This argument at high temperatures is valid at any $U(1)$ phase $\phi$ in $\Omega$, but we would like to comment that the $(\mathbb{Z}_N)_{\mathrm{shift},\rmcq}$ symmetry is enhanced 
to $\mathbb{Z}_N\rtimes \mathbb{Z}_2$ when $\phi$ is quantized to $\pi/N$ although this does not give new anomalies. 
When the flavor-independent boundary condition is set instead of the flavor-twisted boundary condition, no center symmetry exists at generic $\phi$ but a $\mathbb{Z}_2$ subgroup of $\mathbb{Z}_N\rtimes \mathbb{Z}_2$ becomes a symmetry when $\phi\in(\pi/N)\mathbb{Z}$, which  
derives a mixed 't Hooft anomaly with other symmetries when local counterterms cannot cancel it~\cite{Shimizu:2017asf}. This gives an underlying reason for 
Roberge-Weiss phase transition~\cite{Roberge:1986mm}. 
In our case, SSB occurs as $\mathbb{Z}_N\rtimes \mathbb{Z}_2\to \mathbb{Z}_2$, which is consistent with the absence of new anomalies. 

\begin{figure*}[t]
\centering
\begin{minipage}{0.49\textwidth}
\includegraphics[scale=.32]{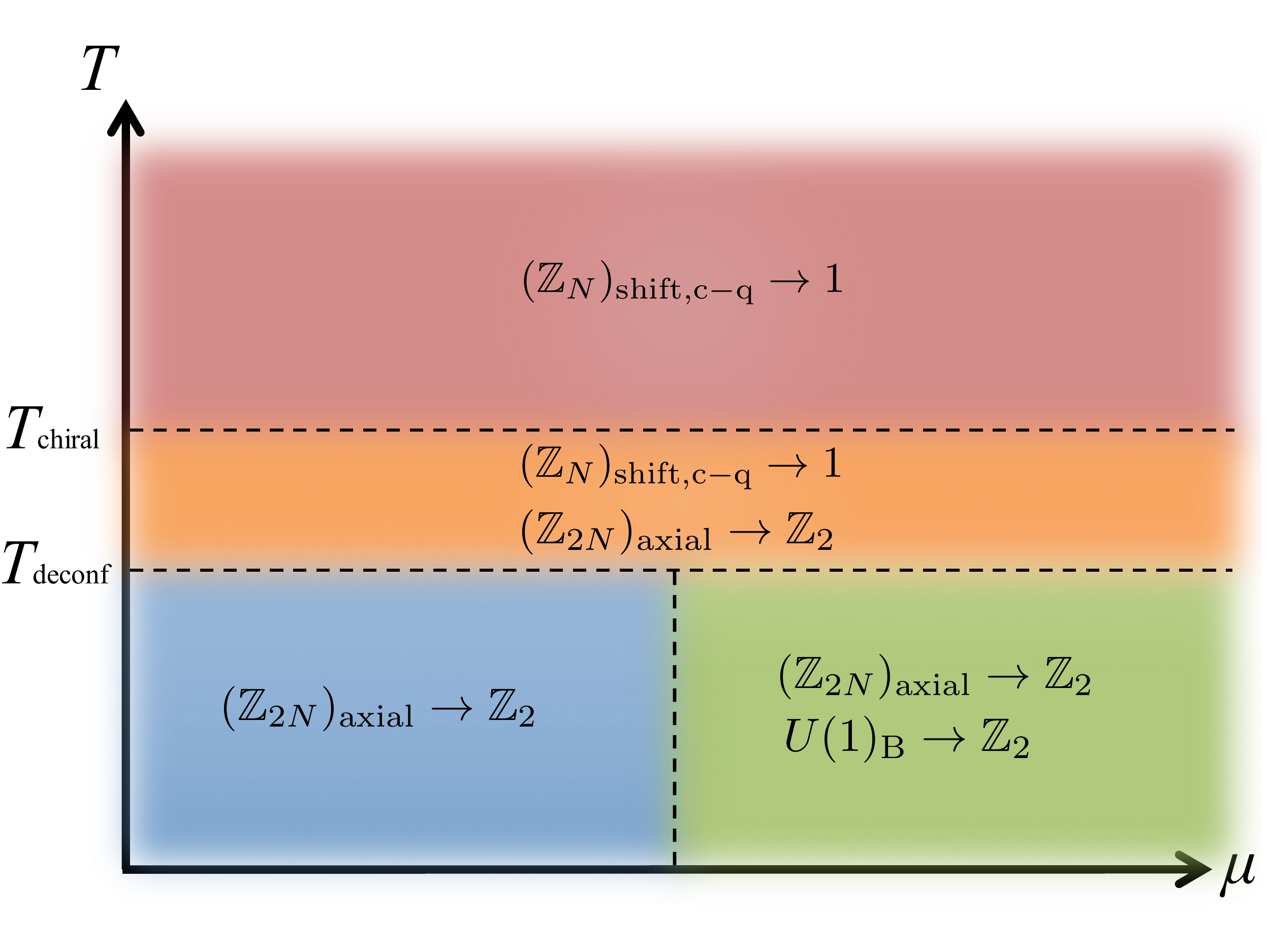}
\end{minipage}
\begin{minipage}{0.49\textwidth}
\includegraphics[scale=.3]{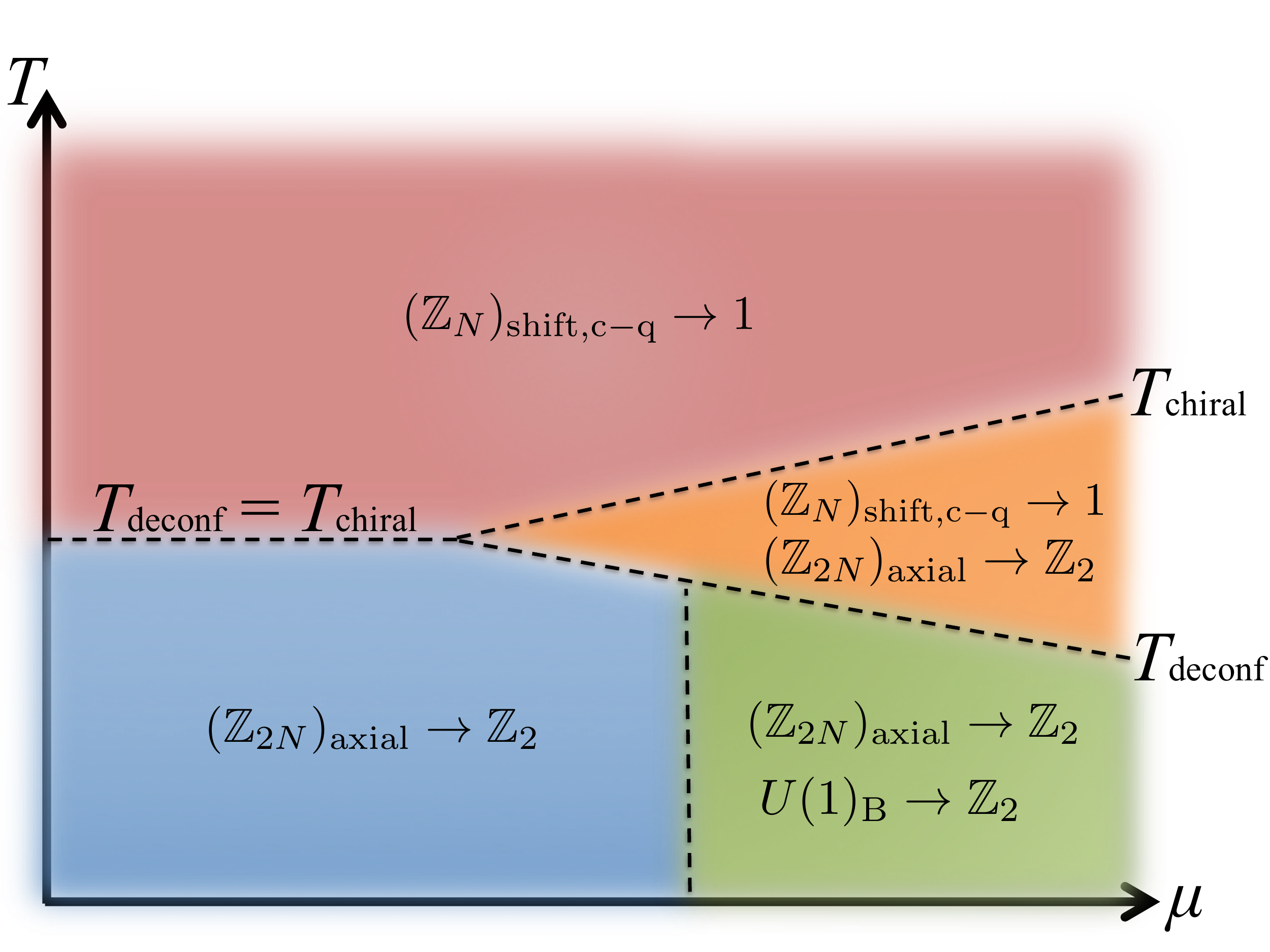}
\end{minipage}
\caption{Two examples of schematic $(T,\mu)$ phase diagrams of massless $\mathbb{Z}_N$-QCD ($N=3$) that are consistent with anomaly. Dashed lines show the phase transition lines, and corresponding symmetry breaking pattern in each phase is specified. $T_{\mathrm{chiral}}$ and $T_{\mathrm{deconf}}$ denote the chiral restoration and deconfinement temperatures, respectively. }
\label{fig:global}
\end{figure*}

\paragraph{High density limit}
In a large-$\mu$ limit, it is widely expected that color superconductivity appears~\cite{Alford:2007xm}. In the following discussion, we set $N=3$. To discuss color superconductivity, we first prepare gauge ``non''-invariant scalar and pseudo-scalar diquark operators ($N=3$), 
\bea
\Delta_{c_1f_1}&=&\ve_{c_1c_2c_3}\ve_{f_1f_2f_3}(q^t_{f_2 c_2}\im \gamma_0\gamma_2 \gamma_5 q_{c_3 f_3}),\nonumber\\
\Delta'_{c_1f_1}&=&\ve_{c_1c_2c_3}\ve_{f_1f_2f_3}(q^t_{f_2 c_2}\im \gamma_0\gamma_2 q_{c_3 f_3}). 
\eea
Since these are not gauge-invariant, this analysis makes sense only at weak coupling but they are useful for Higgs phenomena. 
$(\mathbb{Z}_{2N})_{\mathrm{axial}}$ rotates $\Delta$ and $\Delta'$ by an angle 
$2\pi/N$, and thus it is completely broken down to $\mathbb{Z}_2$ if $\Delta$ or $\Delta'$ gets expectation value. Below, we only use the scalar condensate $\Delta$ via an appropriate rotation of $(\mathbb{Z}_{2N})_{\mathrm{axial}}$.

At sufficiently large $\mu$ and small temperatures, we expect to have $\Delta_{cf}=\Delta_{\mathrm{CFL}} \delta_{cf}$, which is called color-flavor locking (CFL)~\cite{Alford:2007xm}. 
In this phase, the Higgs phenomenon with fundamental scalars occurs and all gluons become massive, which is why weak-coupling analysis is applicable. 
Let us combine it with the analysis of anomaly matching in order to speculate physics in the strongly-coupled regime. 
For that purpose, it is important to identify the pattern of SSB by gauge-invariant order parameters~\cite{Cherman:2017tey}: $\Delta^{\dagger}\Delta\sim (\overline{q}\overline{q})qq$ and $\ve_{c_1c_2c_3}\Delta_{c_1f_1}\Delta_{c_2f_2}\Delta_{c_3f_2}\sim (qq)^3$. 
The latter clearly breaks $U(1)_{\mathrm{B}}\equiv U(1)_\rmq/(\mathbb{Z}_N)_{\rmcq}\to \mathbb{Z}_2$, which is the superfluid phase. 
As we have already discussed, the diquark condensate also breaks $(\mathbb{Z}_{2N})_{\mathrm{axial}}$. Indeed, this is required by anomaly matching because we can obtain the anomaly (\ref{eq:anomaly_massless4d_qcd}) without gauging $U(1)_{\rmq}$ symmetry by setting $B_\rmc=B_\rmf$ when $N\ge 3$~\cite{Tanizaki:2017qhf}. 
In CFL phase, $(\mathbb{Z}_N)_{\mathrm{shift},\rmcq}$ is unbroken. 
We remark that the anomaly is also consistent with another possibility of CFL-like phase, 
where three diquark condensates have distinct values and $(\mathbb{Z}_N)_{\mathrm{shift},\rmcq}$ is broken~\cite{Kouno:2015sja}.

It is also expected from the analysis of effective models~\cite{Kouno:2015sja} that the two-flavor pairing phase (2SC) or the u-flavor pairing phase (uSC) exist at large-$\mu$ regions. 
These phases are characterized by $\Delta_{cf}=\Delta_{2\mathrm{SC}}\delta_{c,3}\delta_{f,3}$ and by $\Delta_{cf}=\Delta_{\mathrm{uSC}}(\delta_{c,2}\delta_{f,2}+\delta_{c,3}\delta_{f,3})$, respectively. 
In uSC phase, all gluons become massive, while there are $SU(2)$ massless gluons in 2SC phase and perturbative computations are useless in 2SC phase.
In these phases, the $(\mathbb{Z}_N)_{\mathrm{shift},\rmcq}$ is spontaneously broken due to the asymmetric diquark condensates in addition to the breaking of $(\mathbb{Z}_{2N})_{\mathrm{axial}}$, 
while $U(1)_\mathrm{B}$ is unbroken\footnote{In the gauge non-invariant argument, it corresponds to the simultaneous baryon and color rotation $U(1)_{{\mathrm B}+\mathrm{c}}$ in 2SC phase, but such description is ill-defined beyond weak-coupling regimes.} due to $(qq)^3=0$. 
This is again consistent with anomaly matching.

\paragraph{Phase structure of massless $\mathbb{Z}_3$-QCD}
In the known phases discussed above, $U(1)_\rmf^{N-1}$ is always unbroken. From the viewpoint of anomaly matching, there is no reason to rule out SSB of $U(1)_\rmf^{N-1}$. 
Assuming that $U(1)_\rmf^{N-1}$ is unbroken everywhere, we can derive an inequality between the ``deconfinement'' temperature $T_{\mathrm{deconf}}$, above which $(\mathbb{Z}_N)_{\mathrm{shift},\rmcq}$ is broken, and the chiral restoration temperature $T_{\mathrm{chiral}}$, above which $(\mathbb{Z}_{2N})_{\mathrm{axial}}$ is unbroken~\cite{Gaiotto:2017yup, Komargodski:2017smk, Shimizu:2017asf}: When $N\ge 3$, for any $\mu$
\be
T_{\mathrm{chiral}}(\mu) \ge T_{\mathrm{deconf}}(\mu).
\label{eq:inequality_temperatures}
\ee 
Under this assumption, the anomaly matching requires that the discrete axial symmetry $(\mathbb{Z}_{2N})_{\mathrm{axial}}$ must be spontaneously broken at low temperatures, which automatically requires SSB of continuous axial symmetry since $(\mathbb{Z}_{2N})_{\mathrm{axial}}\subset SU(N)_{\mathrm{left}}\times SU(N)_{\mathrm{right}}\times U(1)_{\rmq}$. 

We schematically show finite-$(T,\mu)$ phase diagrams of massless ${\mathbb Z}_{3}$-QCD in Fig.~\ref{fig:global}. 
In the strongly-coupled region, not much is known about massless $\mathbb{Z}_N$-QCD, and we present just two possible scenarios that are consistent with anomaly. 
It is an interesting and testable question whether the inequality (\ref{eq:inequality_temperatures}) at $\mu=0$ is saturated or not for lattice QCD simulation. 
It is notable that the symmetric and gapped phase cannot appear, although such phases were observed in the study based on the effective model~\cite{Kouno:2015sja}.
The appearance of forbidden phases is generic when Ginzburg-Landau-type effective approach is used for QFT with an 't Hooft anomaly. 

\section{Conclusion and Discussion}

In this paper, we for the first time put the anomaly-matching constraint related to center and $(\mathbb{Z}_{2N})_{\mathrm{axial}}$ symmetries on the finite-$(T,\mu)$ phase diagram of the $N$-flavor $SU(N)$ gauge theory under the $\mathbb{Z}_N$ flavor-twisted boundary condition.   
We first derived the mixed 't Hooft anomaly (\ref{eq:anomaly_massless4d_qcd}) of massless $N$-flavor QCD between vector-like continuous symmetry and $(\mathbb{Z}_{2N})_{\mathrm{axial}}$. Using this anomaly, we showed that massless $\mathbb{Z}_N$-QCD also has the 't Hooft anomaly (\ref{eq:tHooft_anomaly_ZnQCD}) among $(\mathbb{Z}_N)_{\mathrm{shift},\rmcq}$, maximal Abelian subgroup of vector-like symmetry, and $(\mathbb{Z}_{2N})_{\mathrm{axial}}$. The anomaly matching argument shows that the phase diagram of massless $\mathbb{Z}_N$-QCD realizes a persistent order, i.e., the trivial phase cannot appear at any temperatures $T$ and any quark chemical potentials $\mu$. 
This is the rigorous result on $N$-flavor massless QCD, and it can be found by considering the \textit{thermal} expectation value of the operator given in (\ref{eq:operator_expression}). 

As a consistency check, we compare this result with the known results about $\mathbb{Z}_N$-QCD. At low temperatures and low densities, the lattice simulation shows that discrete axial symmetry is spontaneously broken by the quark bilinear condensate. At sufficiently high temperatures and low densities, one-loop effective potential of the Polyakov loop breaks the intertwined center symmetry $(\mathbb{Z}_N)_{\mathrm{shift},\rmcq}$, which is also confirmed by the lattice simulation. 
At low temperatures and sufficiently high densities, it is widely believed that the CFL phase appears, which breaks $U(1)_\mathrm{B}$ and $(\mathbb{Z}_{2N})_{\mathrm{axial}}$. 
The 2SC or uSC phases are also observed in the effective model analysis, which breaks $(\mathbb{Z}_N)_{\mathrm{shift},\rmcq}$ as well as 
$(\mathbb{Z}_{2N})_{\mathrm{axial}}$.
In all these cases, SSB occurs so that 't~Hooft anomaly matching is satisfied.  
We further predict that $\mathbb{Z}_N$-QCD must have nontrivial phase, such as SSB, conformal behavior, or topological order, even in the region where the system is strongly coupled and lattice simulation suffers from the sign problem. 
We present two examples of phase diagrams consistent with anomaly matching. 

We would also like to comment some speculative remarks on the $N$-flavor QCD phase diagram with the thermal boundary condition in the zero-temperature limit. 
Since our derivation of anomaly has the four-dimensional origin, the anomaly matching argument is valid however large the size $L$ of compactification is. By taking the zero-temperature limit, we expect that the effect of boundary condition would disappear. If the vector-like flavor symmetry is unbroken, it is indeed reasonable to argue that the effect of flavor-dependence in the boundary condition disappears. Under this assumption, anomaly matching argument claims that finite-density massless $N$-flavor QCD shows nontrivial phase at any quark chemical potentials in the zero-temperature limit.
This result also implies SSB of continuous axial symmetry since $(\mathbb{Z}_{2N})_{\mathrm{axial}}\subset SU(N)_{\mathrm{left}}\times SU(N)_{\mathrm{right}}\times U(1)_{\rmq}$. This gives the phenomenological impact on cold dense $3$-flavor massless QCD, because the possible chiral symmetry breaking is restricted even at finite densities~\cite{Stern:1997ri, Kogan:1998zc}. 
We can check that this is indeed the case at least for small $\mu$ and also for sufficiently large $\mu$ in the zero-temperature limits, where the anomaly is satisfied by SSB of discrete axial symmetry. 
At the zero temperature, anomaly matching for continuous chiral symmetry is also available~\cite{Sannino:2000kg, Hsu:2000by}, and we can obtain further constraints on possible dynamics in cold dense QCD.

\begin{acknowledgments}
T.~M. is grateful to H.~Kouno for the fruitful discussion.
Y.~K. and T.~M. also thank the organizers of the workshop 
``Thermal Quantum Field Theory and Their Applications" 
at Yukawa Institute of Theoretical Physics, Kyoto U. in 28-30 August 2017.
This work is started at the conference ``RIMS-iTHEMS International Workshop on Resurgence Theory'' at Kobe, Japan in 6-8 September 2017, and is completed during the workshop ``Resurgent Asymptotics in Physics and Mathematics'' at Kavli Institute for Theoretical Physics from October 2017. 
The authors greatly appreciate these opportunities and hospitalities of organizers. 
Y.~T. is financially supported by RIKEN special postdoctoral program. 
Y.~K. is supported by the Grants-in-Aid for JSPS fellows (Grant No.15J01626).
This work is supported in part by the Japan Society for the 
Promotion of Science (JSPS) Grant-in-Aid for Scientific Research
(KAKENHI) Grant Numbers 16K17677 (T.~M).
T.~M and N.~S are also supported by MEXT-Supported Program for the Strategic Research Foundation
at Private Universities (Keio University) ``Topological Science" (Grant No. S1511006). 
Research at KITP is supported by the National Science Foundation under
Grant No. NSF PHY-1125915.

While completing the draft of this paper, the authors were notified that a paper on $S^1$-compactified QCD with some related aspects will appear~\cite{Cherman:2017dwt}. The authors thank A.~Cherman and M.~\"{U}nsal for sharing the unpublished manuscript. 
\end{acknowledgments}

\appendix

\bibliography{./QFT,./lefschetz}

\end{document}